# Molecular Collapse States in Elliptical Graphene/WSe$_2$ Heterostructure Quantum Dots

Qi Zheng[1,§], Yu-Chen Zhuang[2,§], Ya-Ning Ren[1], Chao Yan[1], Qing-Feng Sun[2,3,4,†], Lin He[1,†]

[1] Center for Advanced Quantum Studies, Department of Physics, Beijing Normal University, Beijing, 100875, People's Republic of China
[2] International Center for Quantum Materials, School of Physics, Peking University, Beijing, 100871, China
[3] Collaborative Innovation Center of Quantum Matter, Beijing 100871, China
[4] Beijing Academy of Quantum Information Sciences, West Bld. #3, No. 10 Xibeiwang East Road, Haidian District, Beijing 100193, China

[§]These authors contributed equally to this work.
[†]Correspondence and requests for materials should be addressed to Qing-Feng Sun (email: sunqf@pku.edu.cn) and Lin He (e-mail: helin@bnu.edu.cn).

**In relativistic physics, both atomic collapse in heavy nucleus and Hawking radiation in black hole are predicted to occur through Klein tunneling process that couples particles and antiparticles. Recently, atomic collapse states (ACSs) were explicitly realized in graphene because of its relativistic Dirac excitation with large "fine structure constant". However, essential role of the Klein tunneling on the ACSs remains elusive in experiment. Here we systematically study the quasibound states in elliptical graphene quantum dots (GQDs). Bonding and antibonding molecular collapse states formed by two coupled ACSs are observed in the elliptical GQDs. Our experiments, supported by theoretical calculations, indicate that the antibonding state of the ACSs will change into a Klein-tunneling-induced quasibound state, revealing deep connection between the ACSs and the Klein**

**tunneling.**

One of the most well-known prediction by quantum electrodynamics is atomic collapse states (ACSs) that are predicted to form when the charges of the nucleus exceeded 170 [1-3], posing a natural and fundamental bound on the periodic table of elements. The ACSs are quite similar to the Hawking radiation in black hole since that both of them are predicted to occur via Klein tunneling process [4]. In experiment, the realization of the ACSs in real atom remains a challenge due to the requirement of the highly charged nucleus. Graphene's relativistic Dirac excitation with "small" Fermi velocity c/300 (c is the velocity of light in vacuum) makes it an ideal platform to overcome this challenge [5-7] and, recently, the ACSs are realized in graphene with a significantly low critical charge [8-11]. Although realizing the ACSs has achieved great success in graphene, essential role of the Klein tunneling on the formation of the ACSs remains elusive in experiment.

In this Letter, we systematically study the ACSs and the Klein-tunneling-induced quasibound states [12-20] in elliptical graphene/$WSe_2$ quantum dots (QDs). The elliptical $WSe_2$ QDs introduce anisotropic Coulomb-like electrostatic potentials on the massless Dirac fermions and generate two coupled ACSs, which further result in bonding and antibonding molecular collapse states in the elliptical graphene QDs (GQDs). Our result indicates that the antibonding molecular state of the ACSs will change into a Klein-tunneling-induced quasibound state, revealing deep connection between the ACSs and the Klein tunneling.

In our experiment, the graphene/WSe$_2$ heterostructure was obtained by a wet transfer fabrication of a graphene monolayer on freshly exfoliated thick WSe$_2$ sheets (see supplementary materials for details [21]). As reported in our previous work, nanoscale WSe$_2$ QDs were frequently observed at the interface of the graphene/WSe$_2$ heterostructure [11]. Figure 1(a) shows a representative scanning tunneling microscope (STM) image of an elliptical graphene/WSe$_2$ heterostructure QD (see Fig. S1 for more experimental data [21]). The thickness of the elliptical QD is 0.8 nm, the same as that of a WSe$_2$ monolayer. The major and minor diameter of the elliptical QD are about 24.7 nm and 17.2 nm, respectively. Figure 1(b) shows atomic-resolved STM image of the elliptical graphene/WSe$_2$ heterostructure QD and its Fast Fourier transform (FFT) image. The relative rotation angle between the WSe$_2$ and graphene is about 26° and there is no signal of atomic defects and strain structure in graphene around the elliptical QD. The dangling bonds at the edge of the circular WSe$_2$ QDs can introduce Coulomb-like electrostatic potentials on the massless Dirac fermions of graphene above them, which generate both the Klein-tunneling-induced quasibound states, *i.e.*, whispering gallery mode (WGM), and the ACSs in the circular GQDs [11]. In this work, we focus on elliptical graphene/WSe$_2$ heterostructure QDs. In a two-dimensional isolated elliptical conductor, the charge density in equilibrium will exhibit a pronounced anisotropy, which is quite different from the uniform charge density in the circular conductor (see Fig. 1(c) and Fig. S2 [21]). Such an anisotropic feature of the potential is expected to strongly affect the confined states in the elliptical GQDs.

Figure 1(d) shows three representative scanning tunneling spectroscopy (STS), *i.e.*, d*I*/d*V*, spectra measured around the elliptical GQD. A typical V-shaped spectrum can be observed in graphene off the elliptical GQD with the Dirac point at $E_D^{Off} \approx 0.12$ eV. Whereas, for the d*I*/d*V* spectra measured on the elliptical GQD, the Dirac point is shifted to $E_D^{On} \approx 0.4$ eV and a sequence of resonance peaks, which arise from the temporarily confined quasibound states [8-20], are clearly observed. To further explore electronic properties of the elliptical GQD, we performed the radially d*I*/d*V* spectroscopic maps along the major (Y) and minor (X) axes, as shown in Fig. 1(e) (see Figs. S3 and S4 for the radially d*I*/d*V* spectroscopic maps in more elliptical GQDs [21]). Significant difference of electrostatic potentials along the two directions can be observed. According to the spatial dependence of global local density of states (LDOS) in the maps [11], the confined potential along both the X and Y axes around the elliptical GQDs can be described by a Coulomb-like electrostatic potential:

$$V_{x/y}(r) = \begin{cases} \hbar v_F \frac{\beta_{x/y}}{d_{x/y}}, & r \leq d_{x/y} \\ \hbar v_F \frac{\beta_{x/y}}{r}, & r > d_{x/y} \end{cases} \quad (1),$$

where $\hbar$ is the reduced Planck constant, $v_F$ is the Fermi velocity, $r$ is the distance from the center of GQD, $\beta_{x/y} = Z_{x/y}\alpha$ with $Z_{x/y}$ the number of charges and $\alpha \sim 2.5$ the fine structure constant of graphene [6], and $d_{x/y}$ is the cutoff radius of the Coulomb potential. According to the measured electrostatic potentials, we obtain $\beta_y$ = 6, $d_y$ = 7.5 nm in the Y axis and $\beta_x$ = 3.6, $d_x$ = 4.5 nm in the X axis (Fig. 1(e)). Here $\beta_y > \beta_x$ because of the higher charge density on the Y axis (see Fig. 1(c) [21]). The anisotropy of the potentials strongly affects the spatial distribution of the quasibound states and the features of the quasibound states measured in the two directions are quite different (Fig. 1(e)). To explicitly show effects of the anisotropic potential on the quasibound states, we carry out energy-fixed d$I$/d$V$ mappings, which reflect spatial distributions of the LDOS at different energies. Figures 2(a)-2(d) (top panels) show spatial distributions of several representative quasibound states (see Fig. S5 for more energy-fixed d$I$/d$V$ mappings at different energies [21]). Obviously, they are quite different from that of the circular GQDs [11] and exhibit anisotropic features.

To fully understand the electronic properties of the elliptical GQDs, we numerically study the problem with an elliptical Coulomb-like electrostatic potential (see Fig. S6 and supplementary materials for the details):

$$V_e(x,y) = \begin{cases} \frac{\hbar v_f \beta_0}{r_0} = \frac{\hbar v_f \beta_x}{d_x} = \frac{\hbar v_f \beta_y}{d_y} & \sqrt{(x/d_x)^2 + (y/d_y)^2} \leq 1 \\ \frac{\hbar v_f \beta_0/r_0}{\sqrt{(x/d_x)^2+(y/d_y)^2}} & \sqrt{(x/d_x)^2 + (y/d_y)^2} > 1 \end{cases} \quad (2).$$

With considering the parameters $\beta_{x/y}$ and $d_{x/y}$ extracted from the experimental results, we obtain theoretical d$I$/d$V$ spectra (Fig. 1(f)), space-energy maps of the LDOS along the X and Y axes of the elliptical GQD (Fig. 1(g)), and spatial distributions of four representative quasibound states N1-N4 (Figs. 2(a)-2(d), bottom panels). Obviously, they are in good agreement with the experimental results. In order to explore the nature of these quasibound states, we calculate the evolution of the quasibound states from the circular GQD to the elliptical GQD (see Figs. S6 and S7 [21]). For a circular GQD with a cut-off radius $r_0$ of the Coulomb potential, there are both the ACSs and the Klein-tunneling-induced WGM states [11]. The ACSs locate at the center of the GQD and follow an exponential function in the energy levels of the quasibound states. The almost equally energy spaced quasibound states at the edge of the GQD are the Klein-tunneling-induced WGM (see Figs. S6 and S7 [21]). The anisotropic elliptical Coulomb potential with cut-off radii $d_x$ and $d_y$ can be obtained by a smooth transition from a circular Coulomb potential (here $d_x = r_0 - d_r$ and $d_y = r_0 + d_r$, $d_r$ is length of the deformation along the two directions, see Fig. S6 and the details of discussion [21]). Here the N1 and N2 quasibound states in the elliptical GQD evolve from the first ACS and first WGM in the circular GQD, respectively. Additionally, they have larger LDOS on major (Y) axis due to the anisotropy of the elliptical potential. Our experiments, supported by our theoretical calculations, indicate the coexistence of electron WGMs and ACSs in the elliptical GQDs, which respectively locate at the edge and the center of the GQD. In our experiment, we can observe Friedel Oscillations outside the GQD due to the screening of the electrostatic potential away from the GQD [11] (See Fig. S8 for the details of discussion [21]). Such an effect does not affect the quasibound states in the GQD and is not taken into account in our calculations.

Thanks to the large cut-off radii of the elliptical GQD, we can realize both the WGMs and ACSs in a single GQD and simultaneously image them, as shown in Figure 2. The spatial distributions of the WGMs and ACSs in the elliptical GQD are quite different from that in the circular GQD (Fig. S9 [21]). It is interesting to note that the WGMs and the ACSs tend to distribute separately in different regions of the elliptical GQD, which can effectively reduce the Coulomb interaction. This phenomenon reminds us of the formation of bonding and antibonding molecular states of two coupled quasibound states [17,22,23], suggesting that the WGMs (ACSs) arise from antibonding (bonding) molecular states. Such a result is unexpected because it is believed that the WGMs [12,24] and the ACSs [5-7] have distinct underlying origins.

To further explore the relationship between the WGMs and the ACSs, we calculate evolution of quasibound states from a molecular Coulomb potential to the elliptical Coulomb potential. Our results indicate that the ACSs (WGMs) in the elliptical GQD arise from bonding (antibonding) molecular states of two coupled ACSs. As shown in Fig. 3(a), the molecular Coulomb potential $V_m(x,y)$ is composed of two identical circular Coulomb potentials separated by a distance $d_y$ on the Y axis:

$$V_m(x, y) = V_A(x, y) + V_B(x, y) \qquad (3).$$

Here the centers of the two circular potential fields are at $(x, y) = (0, \frac{d_y}{2})$ and $(0, -\frac{d_y}{2})$ (see supplementary materials for the details of discussion [21]). Considering an isolated circular Coulomb-like electrostatic potential, the electron WGMs and ACSs coexist [11]. When two circular GQDs are closely coupled, the bonding and antibonding states can be obtained using the linear combination of orbitals (LCAO) method, similar to the $H_2$ molecule [25-27]. It is well-known that the coupling of two hydrogen atoms, *i.e.*, the H$_2$ molecule, due to the coherent superposition and entanglement, will form a lower-energy bonding state and a higher-energy antibonding state. Here, the ACSs and WGMs, both of which can be described by quantum numbers (*m,n*), can be considered as the electron orbits of a circular GQD (see Fig. S7 for the details of discussion [21]). Then the identical quasibound states in the two circular GQDs couple to form the bonding and antibonding states, which are spatially redistributed [17,22,23], as shown in Fig. 3(c). The redistributed quasibound states in the molecular GQD appear to be very similar to that in the elliptical GQD. To analyze the relation between the quasibound states in the elliptical GQD and the molecular GQD, we use the potential field $V(x,y) = (1-\lambda)V_m(x,y) + \lambda V_e(x,y)$ with adjustable $\lambda$ to achieve a smooth evolution from the molecular potential $V_m(x,y)$ to the elliptical potential $V_e(x,y)$, as shown in Fig. 3(b). Figures 3(c)-(e) display the calculated spatial distribution of the quasibound states in the GQDs with different $\lambda$. At $\lambda = 0$, *i.e.*, the standard molecular GQDs, the first two quasibound states (marked by N1 and N2) contribute to the bonding and antibonding states of the first ACS of a single circular GQD, and exhibit the bonding state with LDOS concentrating on the center and antibonding state with LDOS concentrating on the two ends of the molecular GQDs (Fig. 3(c)). With increasing $\lambda$, the ACS bonding state (N1) of the molecular GQDs gradually evolves into the first ACS of the elliptical GQD, and the ACS antibonding state (N2) evolves into the first WGM of the elliptical GQD. We would like to mention that even though here we focus mainly on the origin of the first ACS and the first WGM in the elliptical GQDs, in fact, higher order ACSs and WGMs experience a similar process. Analogously, the bonding state of WGMs will evolve into the new ACSs, but the anti-bonding states of WGMs and unbonded WGMs will evolve into the new WGMs (See Fig. S10 for the details of discussion [21]).

In summary, we systematically study the ACSs and the Klein-tunneling-induced quasibound states in elliptical GQDs. Bonding and antibonding molecular collapse states formed by two coupled ACSs are observed in the elliptical GQDs. Our results indicate that the antibonding state of the ACSs will change into the Klein-tunneling-induced WGMs. The transition between the WGMs and ACSs reveals the deep connection between them.

**References**

[1] Ya. B. Zeldovich, V. S. Popov, Electronic structure of superheavy atoms. *Sov. Phys. Usp.* **14**, 673 (1972).

[2] W. Greiner, B. Muller, and J. Rafelski, Quantum Electrondynamics of Strong Fields (Springer-Verlag, Berlin, 1985).

[3] J. Reinhardt, W. Greiner, Quantum electrodynamics of strong fields. *Rep. Prog. Phys.* **40**, 219–295 (1977).

[4] M. K. Parikh, F. Wilczek, Hawking radiation as tunneling. *Phys. Rev. Lett.* **85**, 5042 (2000).

[5] A. V. Shytov, M. I. Katsnelson, L. S. Levitov, Vacuum Polarization and Screening of Supercritical Impurities in Graphene. *Phys. Rev. Lett.* **99**, 236801 (2007).

[6] A. V. Shytov, M. I. Katsnelson, L. S. Levitov, Atomic Collapse and Quasi–Rydberg States in Graphene. *Phys. Rev. Lett.* **99**, 246802 (2007).

[7] V. M. Pereira, J. Nilsson, A. H. Castro Neto, Coulomb Impurity Problem in

Graphene. *Phys. Rev. Lett.* **99**, 166802 (2007).

[8] Y. Wang, D. Wong, A. V. Shytov, V. W. Brar, S. Choi, Q. Wu, H.-Z. Tsai, W. Regan, A. Zettl, R. K. Kawakami, S. G. Louie, L. S. Levitov, M. F. Crommie, Observing Atomic Collapse Resonances in Artificial Nuclei on Graphene. *Science*. **340**, 734–737 (2013).

[9] J. Mao, Y. Jiang, D. Moldovan, G. Li, K. Watanabe, T. Taniguchi, M. R. Masir, F. M. Peeters, E. Y. Andrei, Realization of a tunable artificial atom at a supercritically charged vacancy in graphene. *Nature Phys.* **12**, 545–549 (2016).

[10] Y. Jiang, J. Mao, D. Moldovan, M. R. Masir, G. Li, K. Watanabe, T. Taniguchi, F. M. Peeters, E. Y. Andrei, Tuning a circular p–n junction in graphene from quantum confinement to optical guiding. *Nature Nanotech.* **12**, 1045–1049 (2017).

[11] Q. Zheng, Y.-C. Zhuang, Q.-F. Sun, L. He, Coexistence of electron whispering-gallery modes and atomic collapse states in graphene/WSe2 heterostructure quantum dots. *Nat Commun*. **13**, 1597 (2022).

[12] A. V. Shytov, M. S. Rudner, L. S. Levitov, Klein Backscattering and Fabry-Pérot Interference in Graphene Heterojunctions. *Phys. Rev. Lett.* 101, 156804 (2008)

[13] Andrea F. Young and Philip Kim, Quantum interference and Klein tunnelling in graphene heterojunctions. *Nature Phys*. **5**, 222–226 (2009).

[14] Y. Zhao, J. Wyrick, F. D. Natterer, J. F. Rodriguez-Nieva, C. Lewandowski, K. Watanabe, T. Taniguchi, L. S. Levitov, N. B. Zhitenev, J. A. Stroscio, Creating and probing electron whispering-gallery modes in graphene. *Science*. **348**, 672–675 (2015).

[15] C. Gutiérrez, L. Brown, C.-J. Kim, J. Park, A. N. Pasupathy, Klein tunnelling and electron trapping in nanometre-scale graphene quantum dots. *Nature Phys*. **12**, 1069–1075 (2016).

[16] Z.-Q. Fu, K.-K. Bai, Y.-N. Ren, J.-J. Zhou, L. He, Coulomb interaction in quasibound states of graphene quantum dots. *Phys. Rev. B*. **101**, 235310 (2020).

[17] Z.-Q. Fu, Y. Pan, J.-J. Zhou, K.-K. Bai, D.-L. Ma, Y. Zhang, J.-B. Qiao, H. Jiang, H. Liu, L. He, Relativistic Artificial Molecules Realized by Two Coupled Graphene Quantum Dots. *Nano Lett.* **20**, 6738–6743 (2020).

[18] K.-K. Bai, J.-J. Zhou, Y.-C. Wei, J.-B. Qiao, Y.-W. Liu, H.-W. Liu, H. Jiang, L. He, Generating atomically sharp p − n junctions in graphene and testing quantum electron optics on the nanoscale. *Phys. Rev. B*. **97**, 045413 (2018).

[19] J. Lee, D. Wong, J. Velasco Jr, J. F. Rodriguez-Nieva, S. Kahn, H.-Z. Tsai, T.

Taniguchi, K. Watanabe, A. Zettl, F. Wang, L. S. Levitov, M. F. Crommie, Imaging electrostatically confined Dirac fermions in graphene quantum dots. *Nature Phys*. **12**, 1032–1036 (2016).

[20] F. Ghahari, D. Walkup, C. Gutiérrez, J. F. Rodriguez-Nieva, Y. Zhao, J. Wyrick, F. D. Natterer, W. G. Cullen, K. Watanabe, T. Taniguchi, L. S. Levitov, N. B. Zhitenev, J. A. Stroscio, An on/off Berry phase switch in circular graphene resonators. *Science*. **356**, 845–849 (2017).

[21] See Supplemental Materials for more experimental data, analysis, and details of discussion.

[22] R. Van Pottelberge, D. Moldovan, S. P. Milovanović, F. M. Peeters, Molecular collapse in monolayer graphene. *2D Mater.* **6**, 045047 (2019).

[23] J. Lu, et al., Frustrated supercritical collapse in tunable charge arrays on graphene. *Nature Commun*. **10**, 477 (2019).

[24] M. I. Katsnelson, K. S. Novoselov, A. K. Geim, Chiral tunnelling and the Klein paradox in graphene. *Nature Phys*. **2**, 620–625 (2006).

[25] W. Heitler, F. London, Wechselwirkung neutraler Atome und homoopolare Bindung nach der Quantenmechanik. *Z. Physik*. **44**, 455–472 (1927).

[26] J. E. Lennard-Jones, The electronic structure of some diatomic molecules. *Trans Faraday Soc.* **25**, 668 (1929).

[27] R. S. Mulliken, Spectroscopy, Molecular Orbitals, and Chemical Bonding. *Science* **157**, 13 (1967).

**Acknowledgments:**

This work was supported by the National Key R and D Program of China (Grant Nos. 2021YFA1401900, 2021YFA1400100 and 2017YFA0303301) and National Natural Science Foundation of China (Grant Nos. 12141401, 11974050, 11921005).

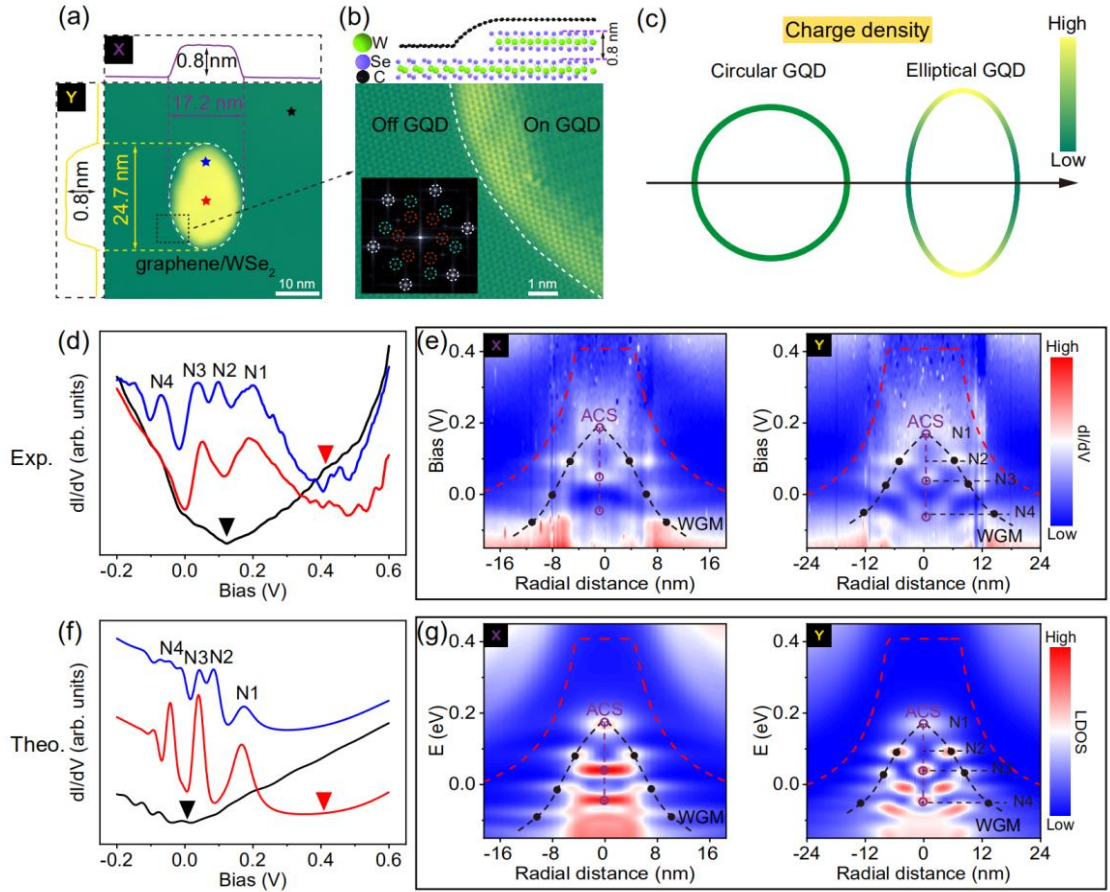

FIG. 1. The structure and electronic properties of an elliptical Graphene/WSe$_2$ heterostructures QD. (a) STM image of a typical elliptical GQD (bias $V$ = 600 mV, set point $I$ = 100 pA), with the height ~0.8 nm. (b) Top panel: Schematic lattice structure of the Graphene/WSe$_2$ heterostructure. Bottom panel: Zoom-in image of the area in black dashed square of panel (a). Inset of bottom panel of (b): The FFT of the Graphene/WSe$_2$ heterostructure. The bright spots in the white dotted circles represent the reciprocal lattice of graphene, the bright spots in the green dotted circles represent the reciprocal lattice of WSe$_2$, and the bright spots in the red dotted circles represent moiré structure of the Graphene/WSe$_2$ heterostructure. (c) Calculated charge density distributions for isolated circular and elliptical conductors, respectively. (d) Typical d$I$/d$V$ spectra taken on and off an elliptical GQD. The measured positions of the colored

spectra are marked with the corresponding color in panel A. (e) The radially d*I*/d*V* spectroscopic maps along the X axis (Left) and Y axis (Right) of the elliptical GQD. (f) The calculated LDOS corresponds to (d). (g) The calculated space-energy maps of the LDOS along the X axis (Left) and Y axis (Right) of the elliptical GQD. The red dashed lines in (e) and (g) indicate Dirac point energy. The black solid dots indicate the quasibound states via the WGMs confinement, and the purple hollow dots indicate ACSs.

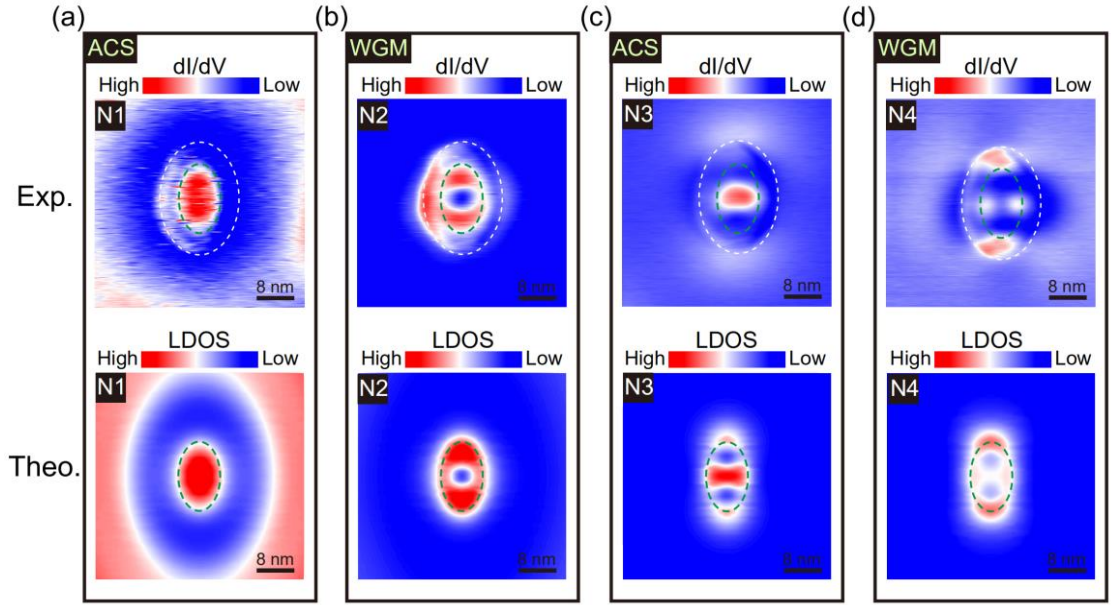

FIG. 2. Spatial distributions of the quasibound states in the elliptical GQD. (a) to (d) Top panels: Energy-fixed d$I$/d$V$ maps at $V_{bias}$ = 0.19 V (N1), at $V_{bias}$ = 0.098 V (N2), at $V_{bias}$ = 0.05 V (N3) and at $V_{bias}$ = -0.05 V (N4) in the elliptical GQD. N1-N4 are the quasibound states (marked in Fig. 1) in the elliptical GQD. Bottom panels: The calculated energy-fixed LDOS in the elliptical GQD at $E$ = 0.19 eV (N1), at $E$ = 0.09 eV (N2), at $E$ = 0.04 eV (N3) and at $E$ = -0.059 eV (N4). The white dashed lines indicate profile of the elliptical GQD. The green dotted lines indicate the cutoff area of the Coulomb-like electrostatic potential in the elliptical GQD.

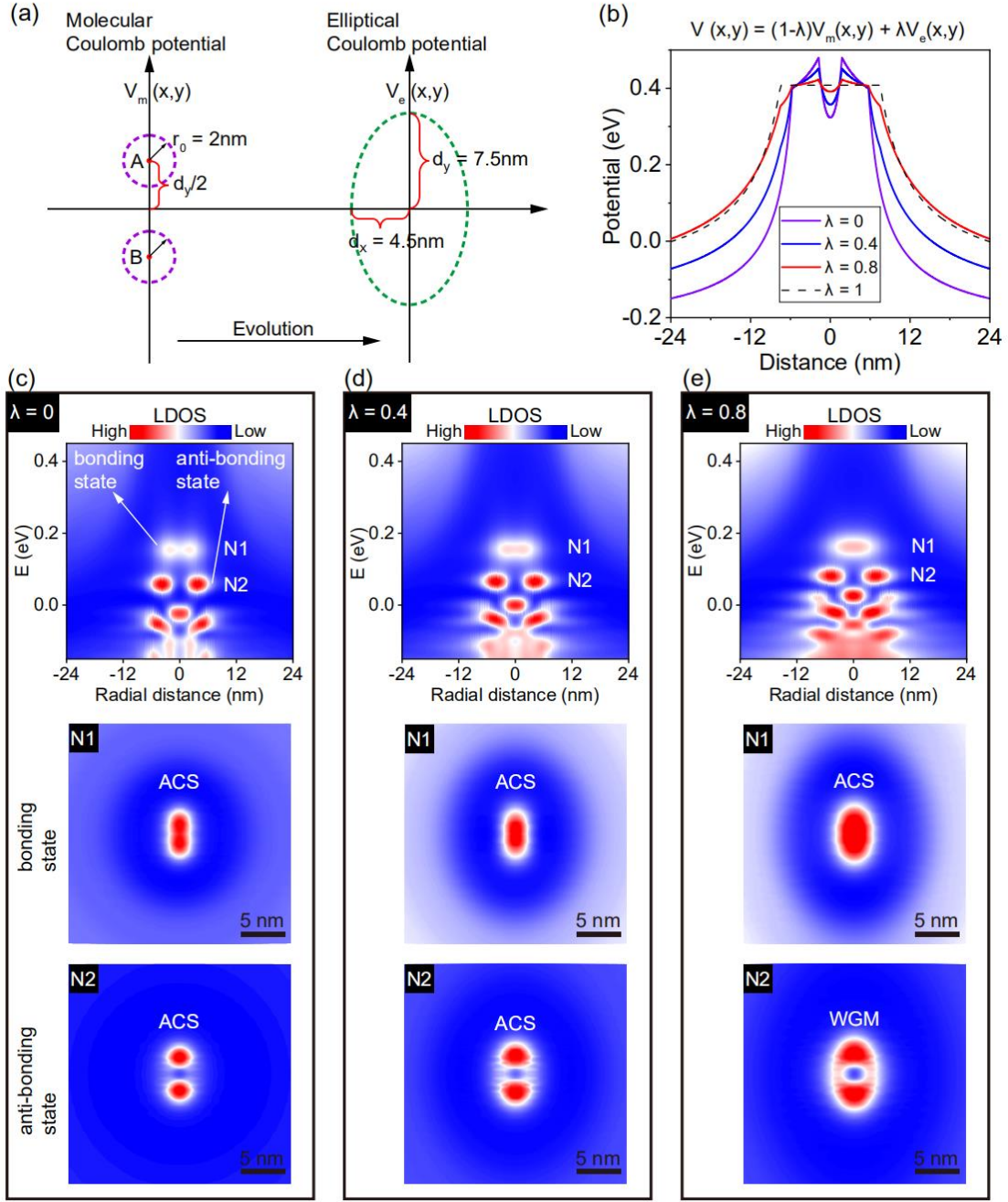

FIG. 3. Evolution from the molecular GQD to the elliptical GQD. (a) The configurations of molecular Coulomb potential and elliptical Coulomb potential. The molecular Coulomb potential consists of two identical circular GQDs (cut-off radius $r_0$ = 2 nm) arranged along the Y axis at a distance of $d_y$ = 7.5 nm. The plateau of the elliptic Coulomb potential is an ellipse with the major axis $d_y$ = 7.5 nm and the minor axis $d_x$ = 4.5 nm. (b) The potential profile of the joint field $V(x,y)$ for different $\lambda$.

(c) to (e) Top panels: The calculated LDOS space-energy maps along Y axis of the GQDs for different $\lambda$. Middle and Bottom panels: The corresponding LDOS maps of the first two quasibound states N1 and N2 for different $\lambda$.